\documentclass[amsmath,amssymb,
aps,
pra,
reprint,
citeautoscript, superscriptaddress,
floatfix]{revtex4-2} 

\usepackage{bbm}
\usepackage{mathrsfs}
\usepackage{bm}
\usepackage{gensymb}
\usepackage{amssymb}
\usepackage{amsfonts}
\usepackage{cancel}
\usepackage{mathtools}
\usepackage{tabularx}

\usepackage[singlelinecheck=false,justification=raggedright]{caption}
\usepackage[singlelinecheck=false,justification=raggedright]{subcaption}
\usepackage{graphicx}  
\usepackage{dcolumn}
\usepackage{placeins}
\usepackage{enumitem}
\usepackage[colorlinks,linkcolor=red,anchorcolor=green,
citecolor=blue,breaklinks]{hyperref}

\usepackage{multirow}
\usepackage{array}

\usepackage{color}
\usepackage{braket}
\usepackage[normalem]{ulem}

\newenvironment{dataavailability}
    {\par\begin{center}\textbf{\MakeUppercase{Data Availability}}\end{center}\noindent}
    {\par}



\usepackage{xcolor}
\usepackage{physics}
\usepackage[outline]{contour} 

\usepackage{tikz}
\usepackage{circuitikz}
\usepackage{quantikz}
\usepackage{pgfgantt}
\usepackage{siunitx}
\usepackage{tikz-3dplot}

\usepackage{pgfplots}
\pgfplotsset{compat=newest}

\usetikzlibrary{3d}
\usetikzlibrary{angles,quotes} 
\usetikzlibrary{patterns} 
\usetikzlibrary{backgrounds}
\usetikzlibrary{snakes}
\usetikzlibrary{fit}
\usetikzlibrary{calc, decorations.pathreplacing, decorations.markings} 
\usetikzlibrary{decorations.pathmorphing}

\tikzset{>=latex} 

\usetikzlibrary{fit, positioning}
\usetikzlibrary{intersections, patterns, pgfplots.fillbetween}
\usetikzlibrary{arrows, arrows.meta, shapes.geometric}

\usetikzlibrary{hobby} 

\definecolor{tensorblue}{rgb}{0.8,0.8,1}
\definecolor{tensorred}{rgb}{1,0.5,0.5}
\definecolor{tensorpurp}{rgb}{1,0.5,1}

\tikzset{ten/.style={fill=tensorblue}}
\tikzset{tenred/.style={fill=tensorred}}
\tikzset{tengreen/.style={fill=green!50!black!50}}
\tikzset{tenpurp/.style={fill=tensorpurp}}
\tikzset{tengrey/.style={fill=black!20}}
\tikzset{tenorange/.style={fill=orange!30}}
\tikzset{u/.style={fill=blue!20,draw=black}}
\tikzset{w/.style={fill=green!50!black!80,draw=black}}

\pgfdeclarelayer{l1}
\pgfdeclarelayer{l2}
\pgfsetlayers{l1,l2,main}

\definecolor{myred}{HTML}{FF1F5B}
\definecolor{myblue}{HTML}{009ADE}
\definecolor{mygreen}{HTML}{00CD6C}
\definecolor{myorange}{HTML}{F27522}
\colorlet{mucol}{red!90!black}
\colorlet{agg_col}{blue!80!}

\tikzstyle arrowstyle=[scale=1]
\tikzstyle directed=[postaction={decorate,decoration={markings,
    mark=at position .65 with {\arrow[arrowstyle]{stealth}}}}]
\tikzstyle reverse directed=[postaction={decorate,decoration={markings,
    mark=at position .65 with {\arrowreversed[arrowstyle]{stealth};}}}]

\tikzstyle{vector_red}=[->,line width=2.0pt,mucol]
\tikzstyle{vector_blue}=[->,line width=2.0pt,myblue]
\tikzstyle{dashed_red}=[->,line width=2.0pt, mucol, dashed]

\makeatletter
\pgfkeys{%
  /tikz/node on layer/.code={
    \gdef\node@@on@layer{%
      \setbox\tikz@tempbox=\hbox\bgroup\pgfonlayer{#1}\unhbox\tikz@tempbox\endpgfonlayer\egroup}
    \aftergroup\node@on@layer
  },
  /tikz/end node on layer/.code={
    \endpgfonlayer\endgroup\endgroup
  }
}

\def\node@on@layer{\aftergroup\node@@on@layer}
\makeatother

\tikzset{%
  >={Latex[width=2mm,length=2mm]},
            base/.style = {rectangle, rounded corners, draw=black,
                           minimum width=4cm, minimum height=1cm,
                           text centered}, 
  activityStarts/.style = {base, fill=blue!20},
       startstop/.style = {base, fill=red!30},
    activityRuns/.style = {base, fill=green!30, minimum size=2mm},
         process/.style = {draw, fill=orange!15, dashed, inner sep = 5pt, 
                           },
  lvl1/.style={draw,fill=red!20,rounded corners=0.2cm,inner sep=5pt,node on layer=l1,  minimum size=2mm},
  lvl2/.style={draw,fill=blue!25,rounded corners=0.2cm,inner sep=5pt,node on layer=l2,  minimum size=2mm},
  lvl3/.style={draw=blue,fill=white,dashed,rounded corners=0.25cm,align=flush center,text width=12em,inner sep=4pt,minimum height=1.5cm}, 
  title/.style={node font=\large,  minimum size=1mm},
  myarrow/.style={latex-latex,ultra thick,blue!80},
}

\tikzset{circle with arrow/.style args={#1,#2}{
        circle, draw=none, 
        thick, inner sep=1pt,
        minimum size=12pt, fill=#1,
        font=\sffamily\bfseries\color{#2}
    },
    arrow style/.style={
        ->, line width=1.4pt, 
        >=latex,
        white
    }
}

\tikzset{
  panel label/.style={
    anchor=north west,
    font=\footnotesize,
    fill=white,
    inner sep=1.5pt,
    rounded corners=1pt
  }
}

\graphicspath{{Figures/}}
\newcolumntype{C}[1]{>{\centering\arraybackslash}p{#1}}

\pdfoutput=1

\begin{document}

\title{Thermalization Regimes in a Chaotic Tavis-Cummings Model}

\author{Sameer Dambal}
\email{sadambal@central.uh.edu}
\affiliation{Department of Physics, 
University of Houston, 
Houston, Texas 77204, United~States}

\author{Eric~R.~Bittner}
\email{ebittner@central.uh.edu}
\affiliation{Department of Physics, University of Houston, Houston, Texas 77204, United~States}

\begin{abstract}
This work investigates the emergent thermalization regimes in a chaotic Tavis-Cummings (TC) model and their implications in quantum spectroscopy. While the TC model is a cornerstone of cavity quantum electrodynamics, traditional treatments often overlook many-body effects that arise in the thermodynamic limit. We utilize the Eigenstate Thermalization Hypothesis to demonstrate that a non-integrable excitonic Hamiltonian within the material manifold drives local thermalization. By tuning the polariton splitting $g$, we observe two dynamical regimes: a thermalizing regime at low interactions driven by quantum chaos and ergodicity, and a non-thermalizing regime at high interactions where strong coupling suppresses ergodicity. We further show that these regimes have direct implications on output photon statistics, specifically influencing the correlation times $\tau_c$ of the cavity population and the second-order correlation function $g^{(2)}(t+\tau)$. We propose that entangled-biphoton spectroscopy serves as an ideal experimental platform to probe these effects and to allow the characterization of the underlying many-body exciton-coupling disorder $\sigma$ through coincidence measurements of the output. Taken together, these results exploit a naturally occurring many-body phenomenon to bridge theoretical predictions with experimental observables.
\end{abstract}

\date{\today}
 
\maketitle

\section{Introduction} \label{sec:introduction}

Optical microcavities interacting with quantum light serve as one of the most versatile platforms for studying complex many-body phenomena, exploring ultrafast physics \cite{reber2016cavity,krischek2010ultraviolet}, probing nonequilibrium systems \cite{nissen2012nonequilibrium, kosior2023nonequilibrium, bakhtiari2015nonequilibrium}, understanding emergent behavior \cite{cho2008fractional}, and constructing large-scale quantum networks \cite{reiserer2015cavity}. High-intensity quantized electromagnetic fields trapped within microcavities allow for experimental tunability at the single-photon level and facilitate the precise manipulation of quantum states, bridging theoretical studies and experimental efforts to explore complex material quantum dynamics and scaling quantum information architectures.

Two prototypical models used in this field are the Dicke model \cite{garraway2011dicke} and the Tavis–Cummings (TC) \cite{bogoliubov1996exact} model. The TC model is exactly solvable \cite{tavis1968exact} under certain conditions and exhibits a plethora of rich physics such as superradiance \cite{dong2022dynamics, chuang2022tavis}, quasiergodic steady states \cite{mondal2024emergence}, Mott insulator to superfluid phase transitions \cite{knap2010quantum, zou2013quantum}, bright and dark states \cite{igorevich2019space}, and other effects emerging from generalized TC models \cite{sun2022dynamics, campos2021generalization}. Recent studies \cite{ray2024ergodic, mondal2025dissipative, torres2017measurement, campos2021generalization, campos2022generalization, agarwal2012tavis, bogoliubov2017time} also show the emergence of quantum chaos in open system TC models, further strengthening the rich physics that this model continues to harbor. As a result, the TC model serves as a central tool for exploring many-body dynamics in materials inside optical microcavities.

While significant progress has advanced our understanding of the TC model, most studies do not capture effects such as localized thermalization and nonintegrability-driven quantum chaos that emerge in the thermodynamic limit. In particular, a growing literature in quantum statistical mechanics suggests that while global information is strictly preserved under Schrödinger evolution of an isolated quantum system, it becomes locally hidden as we scale the system size. This indicates that the entropy of the subsystems increases with time and attains a maximal value \cite{popescu2006entanglement} while preserving the energy of the initial global state. The emergence of such maximal entropy states under these conditions marks the thermalization of the underlying subsystems. In these cases, local observables are effectively described by thermal ensembles despite global preservation of information. These ideas are formally discussed within the framework of the Eigenstate Thermalization Hypothesis (ETH) \cite{deutsch2018eigenstate, srednicki1994chaos}. Thus, it becomes imperative to study the implications of this emergent behavior within analytical frameworks and experiments that extensively utilize the TC model. One such prominent example is quantum spectroscopy \cite{mukamel2020roadmap, dorfman2016nonlinear}.

In this technique, experiments use photons to probe many-body dynamics and correlations in interacting systems by looking at their spectral response. Using the TC model, the spectral response reflects not only individual atomic transitions but also the underlying couplings and interactions between material degrees of freedom. Most molecules and chemical systems studied in quantum spectroscopy exhibit nontrivial couplings and interaction structures that render the underlying material Hamiltonian non-integrable. This non-integrability serves as a prerequisite to ETH-based thermalization within the excitonic manifold and, as a result, strongly shapes the observed spectral response. In this work, we show that these effects leave imprints on the output photon statistics.

We probe these imprints using an emerging technique in quantum spectroscopy known as entangled-biphoton spectroscopy (EBS) \cite{dambal2025quantum, piryatinski2026scattering, schlawin2018entangled, bittner2020probing, li2018probing, li2019photon, eshun2022entangled, yabushita2004spectroscopy, moretti2023measurement, schlawin2013two, roslyak2009nonlinear, malatesta2023optical}. This approach harnesses the entanglement of biphotons to serve as sensitive probes of many-body correlations in complex materials. By allowing these photons to interact with the material placed within a cavity and by observing the change in their entanglement structures in the output statistics, this technique reveals complex correlations at the single excitation level. In this work, we show that in regimes where the system thermalizes under ETH, local observables such as single-exciton and cavity populations attain a steady state at long times. However, in regimes that prevent thermalization, these observables do not attain a steady state. Consequently, the output photon statistics in EBS inherit this behavior and reveal the complex exciton-exciton couplings of the underlying material system.

The central focus of this work is two-fold: (i) We exploit this inevitable localized thermalization of excitons and numerically demonstrate the different thermalization regimes that are consequently manifest in a chaotic TC model; and (ii) we show that these regimes leave imprints on the correlation times of the output photon statistics that can be observed using $g^{(2)}(t)$ measurements in EBS. Our results bridge theory and experiments in a unique manner: By \textit{tuning the cavity-material coupling strength $g$} and by \textit{observing the time-correlation statistics $\tau_c$} of the output biphotons, we leverage the emergent thermalization regimes and extract the underlying \textit{exciton-exciton coupling disorder $\sigma$}.

The study of different forms of disorder in TC models and their implications has garnered wide attention \cite{botzung2020dark, engelhardt2023polariton, wierzchucka2024integrability}. However, most of these studies introduce a disorder in exciton splittings that does not exhibit behaviors that are emergent in the thermodynamic limit. Here, we instead introduce a disorder in the exciton–exciton couplings, which induces quantum chaos and pushes the system to display localized thermalization. We therefore refer to our system as a \textit{chaotic TC model}. In particular, \citet{botzung2020dark} analyze the semilocalization of dark states across multiple noncontiguous sites, while \citet{engelhardt2023polariton} investigate both localized and delocalized regimes, as well as the transition from diffusive to ballistic transport in the same model. By introducing an exciton-exciton coupling disorder $\sigma$ in our work, we reinterpret this localization as ETH-driven localized thermalization. Since ETH applies to eigenstates at finite energy densities, the dark states in a chaotic TC model naturally exhibit these features. We further discuss the limits where these properties vanish in the chaotic TC model. 

In a similar vein, \citet{mattiotti2024multifractality} show how an integrability-breaking term turns nonergodic eigenfunctions, lacking thermalization, into ergodic eigenfunctions that display thermalization. In our work, $\sigma$ breaks the same integrability and makes finite energy density eigenfunctions ergodic and exhibit thermalization. By tuning the exciton-cavity coupling $g$, we switch between the ergodic and nonergodic regimes, thus drawing similar results in these different types of disordered TC models. We also go beyond and show how such phenomena leave distinct imprints in quantum spectroscopic measurements. We note that the chaotic region of the Dicke model also realizes similar ETH-based dynamics \cite{villasenor2022chaos} that may reveal new and interesting results in this context. Other recent works \cite{mikheev2025prethermalization} also study signatures of prethermalization in Rydberg arrays placed in cavities, further reinforcing our motivation to understand their implications in quantum spectroscopy.



The rest of the paper is organized as follows: In Section \ref{subsec:hamiltonian}, we introduce our theoretical model and expound on its relevance in our work. In particular, we highlight interesting spectral properties of our Hamiltonian that reflect signatures of quantum chaos and hint towards different regimes of thermalization. In Section \ref{sec:therm_excitons}, we outline the key theoretical concepts behind isolated system thermalization along with some diagnostic metrics. We discuss our results in Section \ref{sec:results} and elaborate on how they impart unique signatures in the output photon statistics. In Section \ref{sec:experimental_utility}, we propose how EBS can experimentally test our theoretical results and elaborate on how this experiment can leverage our theoretical study to reveal the underlying material correlations by measuring the output photon statistics.

\section{Theoretical Model} \label{sec:theory}

\begin{figure*}[t!]
    \centering
     \begin{subfigure}[b]{0.49\textwidth}
        \centering
        \tdplotsetmaincoords{75}{110} 

\begin{tikzpicture}[tdplot_main_coords, scale=2.5] 

    \tikzset{
        networkNode/.style={
            circle, 
            ball color=orange!80, 
            minimum size=8pt,   
            inner sep=0pt,
            fill opacity=0.9    
        }
    }

    \node[networkNode] (T) at (0, 0.2, 0.5) {};
    \node[networkNode] (A) at (0.5, -0.5, 1) {};   
    \node[networkNode] (B) at (-0.5, 0.866, 1) {}; 
    \node[networkNode] (C) at (0, 0.2, 1.63) {}; 
    \node[networkNode] (D) at (1, 0, 0) {}; 
    \node[networkNode] (E) at (-1, 0.5, -0.3) {}; 

    \begin{scope}[thick, <->, >=stealth, red!40]
        
        \draw (T) -- (A);
        \draw (T) -- (B);
        \draw (T) -- (C);
        \draw (T) -- (D);
        \draw (T) -- (E);
    
        \draw (A) -- (B);
        \draw (A) -- (C);
        \draw (A) -- (D);
        \draw (A) -- (E);

        \draw (B) -- (C);
        \draw (B) -- (D);
        \draw (B) -- (E);

        \draw (C) -- (D);
        \draw (C) -- (E);

        \draw (D) -- (E);
    \end{scope}

    \begin{scope}[thick, green!60!black, dotted, >=stealth, bend right=12]
        \draw (T) to (A); \draw (T) to (B); \draw (T) to (C); \draw (T) to (D); \draw (T) to (E);
        \draw (A) to (B); \draw (A) to (C); \draw (A) to (D); \draw (A) to (E);
        \draw (B) to (C); \draw (B) to (D); \draw (B) to (E);
        \draw (C) to (D); \draw (C) to (E);
        \draw (D) to (E);
    \end{scope}

    \matrix [
        draw, 
        fill=white, 
        anchor=north east, 
        at={(0, -1)}, 
        font=\small,
        row sep=0.1cm
    ] {
        \draw[thick, red!40, <->, >=stealth] (0,0) -- (1.5,0); & \node {u}; \\
        \draw[thick, green!60!black, dotted] (0,0) to (1.5,0); & \node {h}; \\
    };

\end{tikzpicture}
        \caption{}
        \label{fig:exciton_network}
    \end{subfigure}
    \hfill
    \begin{subfigure}[b]{0.49\textwidth}
        \centering
        \usetikzlibrary{decorations.pathmorphing, shadings}

\begin{tikzpicture}

    \def\cavityLen{6}
    \def\mirrorHeight{4.8} 
    \def\mirrorThick{1}    
    \def\bendAngle{25}     
    \def\caseBend{20}
    \def\sagOffset{0.6}

    \fill[blue!10, opacity=0.3] 
        (0, \mirrorHeight/2) to[bend right=\caseBend] (\cavityLen, \mirrorHeight/2)
        to[bend left=\bendAngle] (\cavityLen, -\mirrorHeight/2)
        to[bend right=\caseBend] (0, -\mirrorHeight/2)
        to[bend left=\bendAngle] (0, \mirrorHeight/2);

    \begin{scope}[draw=gray!60, thick]
        \draw (0, \mirrorHeight/2) to[bend right=\caseBend] (\cavityLen, \mirrorHeight/2);
        \draw (0, -\mirrorHeight/2) to[bend left=\caseBend] (\cavityLen, -\mirrorHeight/2);
    \end{scope}

    \coordinate (NetCenter) at (\cavityLen/2, 0);

    \begin{scope}
        \tikzset{
            netNode/.style={circle, ball color=orange!80, minimum size=7pt, inner sep=0pt}
        }
        
        \node[netNode] (C) at ([shift={(0, -0.2)}]NetCenter) {};
        \node[netNode] (N1) at ([shift={(-0.2, 1.2)}]NetCenter) {};    
        \node[netNode] (N2) at ([shift={(1.2, 0.6)}]NetCenter) {};  
        \node[netNode] (N3) at ([shift={(1, -1.2)}]NetCenter) {}; 
        \node[netNode] (N4) at ([shift={(-0.9, -1.4)}]NetCenter) {};
        \node[netNode] (N5) at ([shift={(-1.4, 0.4)}]NetCenter) {}; 

        \begin{scope}[thin, <->, >=stealth, red!40]
            \draw (C) -- (N1);
            \draw (C) -- (N2);
            \draw (C) -- (N3);
            \draw (C) -- (N4);
            \draw (C) -- (N5);

            \draw (N1) -- (N2);
            \draw (N1) -- (N3);
            \draw (N1) -- (N4);
            \draw (N1) -- (N5);

            \draw (N2) -- (N3);
            \draw (N2) -- (N4);
            \draw (N2) -- (N5);

            \draw (N3) -- (N4);
            \draw (N3) -- (N5);

            \draw (N4) -- (N5);         
        \end{scope}

        \begin{scope}[thick, green!60!black, dotted, >=stealth, bend right=12]
        \draw (C) to (N1); \draw (C) to (N2); \draw (C) to (N3); \draw (C) to (N4); \draw (C) to (N5);
        \draw (N1) to (N2); \draw (N1) to (N3); \draw (N1) to (N4); \draw (N1) to (N5);
        \draw (N2) to (N3); \draw (N2) to (N4); \draw (N2) to (N5);
        \draw (N3) to (N4); \draw (N3) to (N5);
        \draw (N4) to (N5);
    \end{scope}
        
    \end{scope}


    \node (OuterCircle)[
        draw=black, 
        dotted,              
        thin, 
        circle,              
        inner sep=-12pt,       
        fit=(N1) (N2) (N3) (N4) (N5) 
    ] {};

    \begin{scope}[red, opacity=0.8, thick, dash pattern=on 2pt off 2pt]
        \draw[samples=200, domain=-\sagOffset:\cavityLen+\sagOffset, variable=\x, smooth] 
            plot (\x, {1.7*sin(deg((\x+\sagOffset)*pi/(\cavityLen+2*\sagOffset)))});
        \draw[samples=200, domain=-\sagOffset:\cavityLen+\sagOffset, variable=\x, smooth] 
            plot (\x, {-1.7*sin(deg((\x+\sagOffset)*pi/(\cavityLen+2*\sagOffset)))});
    \end{scope}

    \draw[fill=blue!60, draw=blue!40!black, thick] 
        (0, \mirrorHeight/2) to[bend right=\bendAngle] (0, -\mirrorHeight/2) -- 
        (-\mirrorThick, -\mirrorHeight/2) -- (-\mirrorThick, \mirrorHeight/2) -- cycle;
    \draw[blue!60!black, ultra thick] 
        (0, \mirrorHeight/2) to[bend right=\bendAngle] (0, -\mirrorHeight/2);

    \begin{scope}[xshift=\cavityLen cm]
        \draw[fill=blue!60, draw=blue!40!black, thick] 
            (0, \mirrorHeight/2) to[bend left=\bendAngle] (0, -\mirrorHeight/2) -- 
            (\mirrorThick, -\mirrorHeight/2) -- (\mirrorThick, \mirrorHeight/2) -- cycle;
        \draw[blue!60!black, ultra thick] 
            (0, \mirrorHeight/2) to[bend left=\bendAngle] (0, -\mirrorHeight/2);
    \end{scope}

    \begin{scope}[cyan!80!blue, thick, opacity=0.8]
        \coordinate (LModeStart) at (-\sagOffset, 0);
        \coordinate (RModeStart) at (\cavityLen + \sagOffset, 0);

        \foreach \angle in {120, 180, 240} {
            \draw[decorate, decoration={snake, amplitude=1pt, segment length=4pt}] 
                (LModeStart) -- (OuterCircle.\angle);
        }

        \foreach \angle in {60, 0, 300} {
            \draw[decorate, decoration={snake, amplitude=1pt, segment length=4pt}] 
                (RModeStart) -- (OuterCircle.\angle);
        }

        \node[cyan!80!black, font=\small, fill=white, inner sep=1.5pt, opacity=0.8, text opacity=1.4] at (-0.2, 0.5) {$\textbf{g}$};
        \node[cyan!80!black, font=\small, fill=white, inner sep=1.5pt, opacity=0.8, text opacity=1.4] at (\cavityLen+0.2, 0.5) {$\textbf{g}$};
    \end{scope}
        
\end{tikzpicture}
        \caption{}
        \label{fig:chaotic_tc}
    \end{subfigure}

    \caption{\textbf{Overview of the disordered excitonic system and cavity-mediated thermalization:} (a) shows a network of excitons with a complex coupling structure. The red lines depict exciton-exciton hoppings and the green dotted lines depict biexciton interactions. This makes the excitonic Hamiltonian non-integrable and promotes ETH-based thermalization. (b) shows a TC model where an optical microcavity couples with the complex excitonic network with the coupling constant $g$.}
    \label{fig:combined_results}
\end{figure*}
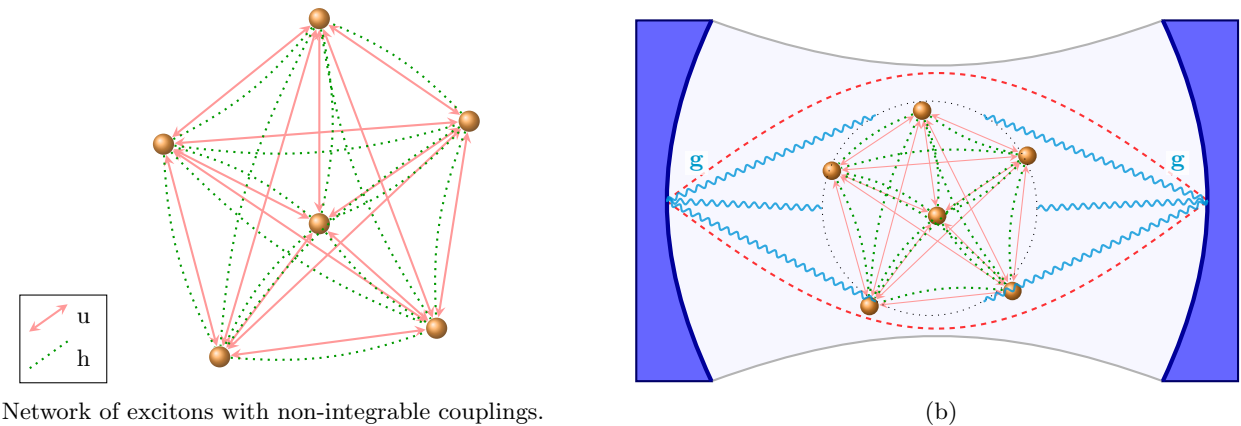

\subsection{Hamiltonian}\label{subsec:hamiltonian}

To interrogate out-of-equilibrium dynamics of excitons and their implications, we study the following Hamiltonian,

\begin{equation}
    \label{full_hamiltonian}
    H = H_c + H_{ex} + H_I 
\end{equation}

where $H_c$ and $H_{ex}$ are the cavity and exciton Hamiltonians, and $H_I$ describes the interaction between the cavity and exciton modes. Specifically, 

\begin{eqnarray}
    \label{cav_hamiltonian}
    H_c = \omega_c a^\dagger a 
\end{eqnarray}

\begin{eqnarray}
    \label{ex_hamiltonian}
    H_{ex} &=& \omega_{ex} \sum_i b_i^\dagger b_i - \sum_{i,j} (h_{ij} b_i^\dagger b_j + h^*_{ji} b^\dagger_j b_i) \nonumber \\ &+& \sum_{kl}u_{kl} n_k n_l
\end{eqnarray}

$a, a^\dagger$ are single-mode bosonic cavity operators obeying $[a,a^\dagger]=1$, and $b_i^\dagger, b_i$ describe hardcore bosonic exciton modes such that $b^2_i = b^{\dagger2}_i = 0$. This enforces at most a single occupancy per site. In addition, $n_k = b^\dagger_k b_k$ is the excitonic number operator that captures the occupation of different excitonic sites. We model the exciton-exciton exchange coupling and biexciton interactions using random variables $h_{ij}, u_{kl}$ that are drawn from two independent and identically distributed Gaussian Unitary Ensembles (GUE) with mean $\langle h\rangle=\langle u\rangle =0$ and variance $\sigma^2 = \frac{1}{d}$. The corresponding standard deviation $\sigma$ describes the \textit{coupling disorder} in excitons and collectively parameterizes $h_{ij}$ and $u_{kl}$. As a result, $\sigma$ captures the collective many-body correlations of materials and therefore becomes a parameter of direct spectroscopic interest in our work. Here, $d$ represents the dimension of the Hilbert space and ensures that the global properties of the ensemble remain normalized as the system scales. The indices $i,j$ and $k,l$ run over all sites in the system.

We adopt this stochastic treatment to account for the structural disorder and complex coupling maps typical of excitonic systems, where a well-structured Hamiltonian is physically unrealistic. Moreover, studies also show that the level spacing statistics of such Hamiltonians, inspired by nuclear spectra, resemble that of the eigenvalues of a random matrix \cite{lopac1996level, wigner1993characteristic}. In the thermodynamic limit, this converges to a Wigner-Dyson (WD) distribution and indicates quantum chaos in the dynamics governed by the underlying Hamiltonian~\cite{d2016quantum}. In the same limit, this promotes subspaces across the system to thermalize~\cite{banuls2011strong} and bridges our theoretical model with realistic phenomena.

To simplify our study, we impose the resonance condition $\omega_c = \omega_{ex} = \omega_0$ and remove detuning as a control parameter in our model. This choice fixes a finite energy offset in the Hamiltonian. Since only the level-spacing statistics and the Hamiltonian’s non-integrability dictate thermalization, the finite offset does not affect the dynamics we study. Finally, the last term in Eq. \eqref{full_hamiltonian} describes the cavity-exciton coupling, 

\begin{equation}
    \label{cav_ex_coupling}
    H_I = g \sum_i (b_i^\dagger a + a^\dagger b_i)
\end{equation}
where $g$ denotes the exchange rate between the cavity and excitons. Taken together, Eq. \eqref{full_hamiltonian} describes a TC model with disordered couplings in the excitonic lattice.

We assume that the system begins with two excitations in the single-mode cavity. One can initialize such a state using an SPDC source that generates entangled biphotons and populates cavity modes \cite{moretti2023measurement, couteau2018spontaneous, wen2007spontaneous}. Since the Hamiltonian $H$ from Eq. \eqref{full_hamiltonian} commutes with the number operator $N = a^\dagger a + \sum_i b_i^\dagger b_i$, it conserves the number of excitations in the system. As a result, the choice of this initial state restricts the dynamics of our model within the two-excitation manifold and simplifies our numerical simulations. We shall see that this initial state forms a typical state in $H$ and consequently represents a large class of initial states that display thermalization.



\subsection{Spectral Statistics}\label{subsec:spectral_statistics}

\begin{figure*}[t!]
    \centering
    \begin{subfigure}[b]{0.49\linewidth}
        \centering
        \includegraphics[width=\linewidth]{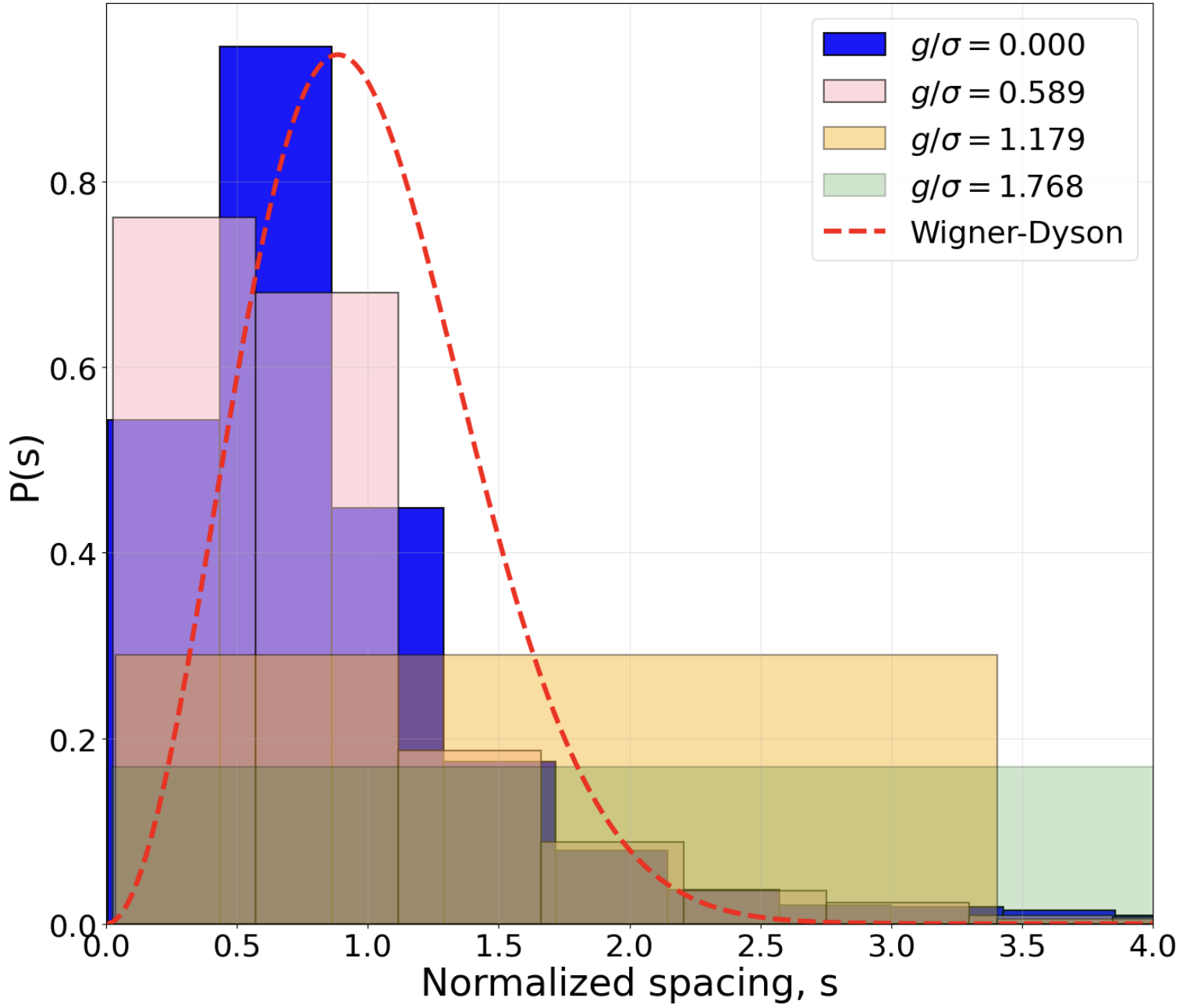}
        \caption{}
        \label{fig:level_spacings}
    \end{subfigure}
    \hfill
    \begin{subfigure}[b]{0.49\linewidth}
        \centering
        \includegraphics[width=\linewidth]{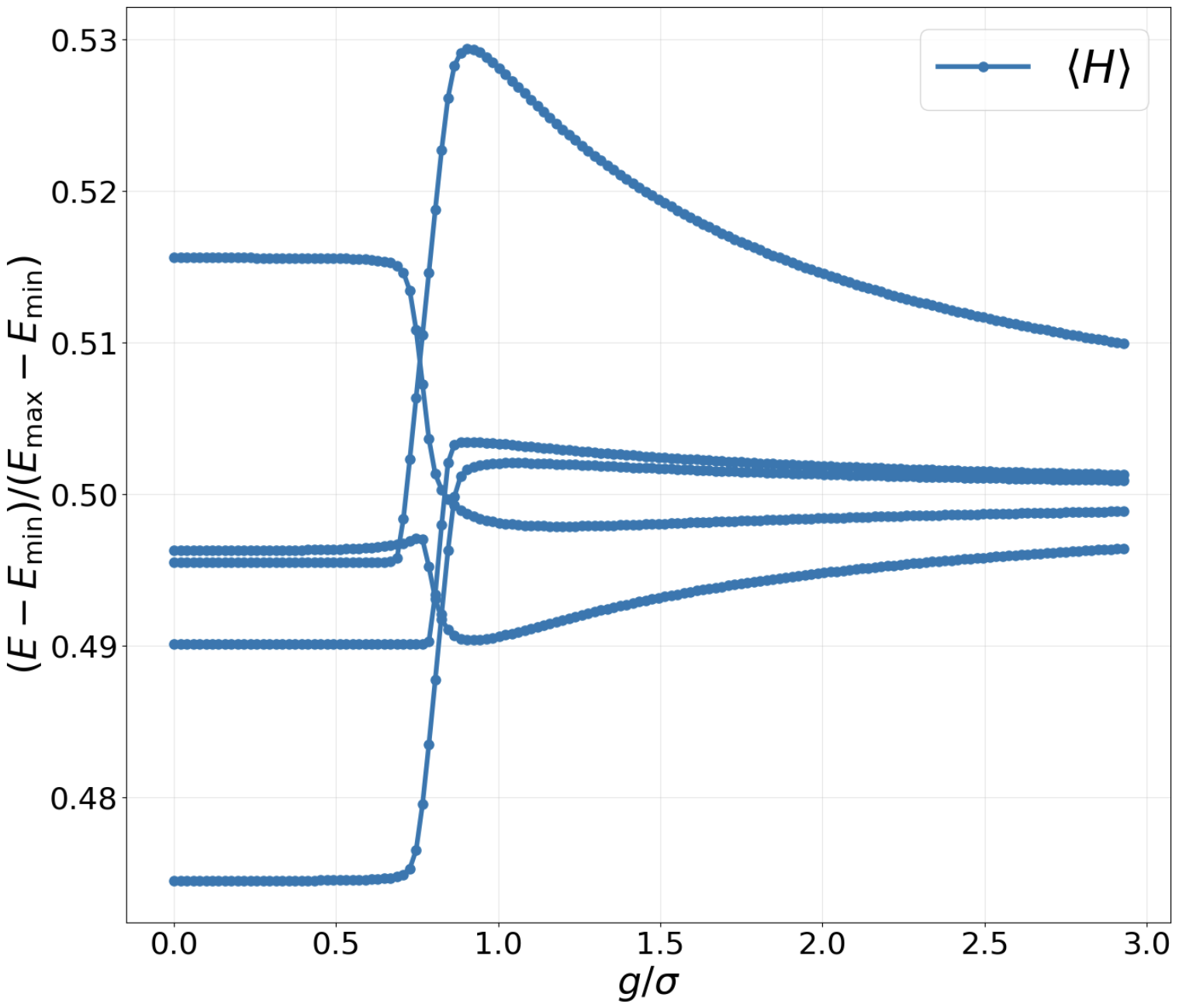}
        \caption{}
        \label{fig:expectation_H}
    \end{subfigure}
    \caption{\textbf{Spectral Statistics:} (a) This describes the level spacing statistics of the eigenvalues of Hamiltonian in Eq. \eqref{full_hamiltonian} projected in the $N \leq 2$ excitation sector. We see that it resembles that of a Wigner-Dyson distribution (dotted red line) for small coupling strengths $g/\sigma < 1$. However, the spectrum flattens for stronger couplings $g/\sigma \gtrsim 1$. These hint towards different regimes of thermalization; (b) This figure shows the normalized energy of the initial state for multiple realizations of the Hamiltonian. Each blue curve corresponds to one realization of $\sigma$ and the random couplings drawn from a GUE. At $g/\sigma <1$, we see the energy of the initial state remains invariant and at $g/\sigma >1$, we see that it saturates towards $0.5$. At the transition point $g/\sigma \sim 1$, we see a sharp change in the energy, which indicates the reorganization of the highest density of states within the spectrum, signaling a change in the dominant underlying quantum dynamics regime.}
    \label{fig:Hamiltonian_statistics}
\end{figure*}

We readily identify two limits in the Hamiltonian $H$ in Eq. \eqref{full_hamiltonian}. For $g \rightarrow 0$, the disordered excitonic component $H_{ex}$ from Eq. \eqref{ex_hamiltonian} dominates $H$, while for $g \rightarrow \infty$, the TC interaction $H_I$ from Eq. \eqref{cav_ex_coupling} dominates $H$. Since $\sigma$ characterizes the strength of the coupling disorder in $H_{ex}$ that governs the spread of excitations within the excitonic manifold, and $g$ controls the oscillation of excitations in and out of this manifold, the parameter $g/\sigma$ captures the dominant regime in this TC model. Naturally, this gives rise to two limits that we label as the \textit{(i) quantum chaos dominant limit, $g/\sigma \rightarrow 0$}, and the \textit{(ii) TC dominant limit, $g/\sigma \gtrsim 1$}. Consequently, the former inherits the WD level spacing statistics characteristic of quantum chaos, and the latter inherits the bright and dark state structures typical in the TC model.

In Figure \ref{fig:level_spacings}, we plot the level spacing statistics for various values of the ratio $g/\sigma$. The dotted red curve shows the WD distribution, 

\begin{eqnarray}
    p_s = \frac{\pi s}{2}e^{-\pi s^2/4}
\end{eqnarray}

where $s$ is the level spacing. At small values of this ratio, we see that the spectrum closely matches the dotted red curve and displays incipient signatures of the WD statistics. At these strengths, quantum chaos predominantly drives the dynamics in the model. At $g/\sigma \gtrsim 1$, we see that the spectrum loses these signatures and flattens. In this limit, the Rabi oscillations induced by the parameter $g$ dictate the dynamics and reduce the Hamiltonian to the TC model. This reflects the increasing dominance of coherent cavity-exciton coupling over random excitonic interactions. 

To make this more concrete, we track the normalized expectation of the Hamiltonian in the initial state $\langle H\rangle = \langle \psi_0 |H|\psi_0\rangle$ as a function of $g/\sigma$ in Figure \ref{fig:expectation_H}. We spectrally expand the initial state in $H$ from Eq. \eqref{full_hamiltonian}, $\ket{\psi_0} = \sum_i \alpha_i \ket{i}$, where $\alpha_i =  \langle i |\psi_0\rangle$ denotes the overlap of the initial state with the eigenstate $\ket{i}$ of $H$. In most many-body systems, the density of states (DoS) is highest near the center of the spectrum. Consequently, the energy of a typical initial state is likely to be located in this region and has the largest overlap $\alpha_i$ for $i$ located around the central region. Stated differently, $\langle H\rangle$ tracks regions with the highest DoS. As we tune the cavity-exciton coupling $g$, these regions shift in the eigenspectrum and $\langle H\rangle$ follows these regions. As a result, the change in the level spacing statistics with $g/\sigma$ influences the locations and distributions of the highest DoS, and reflects any potential regime transitions in $\langle H\rangle$. 

In Figure \ref{fig:expectation_H}, we run the simulation over multiple realizations of the disordered Hamiltonian in Eq. \eqref{full_hamiltonian} for the same initial state of two excitations in the cavity. Each blue curve corresponds to one realization of $\sigma$ and the random couplings drawn from a GUE. We observe that the normalized energies in all realizations remain invariant at low coupling strengths $g$ for all realizations of $H$. This is because the exciton coupling disorder $\sigma$ dominates $H$ and drives the level spacing statistics of all blue curves equally toward a WD distribution. Viewed in the vicinity of $\langle H\rangle$ at $g/\sigma \rightarrow 0$, this leaves the DoS largely unchanged and preserves $\langle H\rangle$. As $g/\sigma \sim 1$, we see sharp transitions in $\langle H\rangle$, suggesting a reorganization of the eigenspectrum and a shift in the location of the highest DoS. This behavior marks a transition between the quantum chaos dominant limit and the TC dominant limit. At $g/\sigma \gtrsim 1$, we observe that the normalized energies $\langle H\rangle$ gradually saturate towards $0.5$. This is because, in the strong coupling limit, the polariton splitting $g\sqrt{N}$ between the upper and lower polaritons becomes large and effectively confines the highest DoS towards the center of the eigenspectrum. As a result, $\langle H\rangle$ tracks this region and saturates to $0.5$ in the large $g$ limit.

This behavior lends similar reasoning to prior studies that examine the role of spectral statistics in the weak-coupling limit of open quantum systems. In particular, \citet{riera2012thermalization} show that the system–environment coupling $V$ and the minimum spectral gap, $\text{min gap}(H)$ are intimately connected. As the dimension of the Hilbert space of the bath increases, the spectrum of a noninteracting Hamiltonian becomes dense and reduces $\text{min gap}(H)$. Consequently, the weak-coupling condition, defined by comparing the interaction energy scale $\lVert V \rVert_\infty$ with $\text{min gap}(H)$, breaks down, leading to memory effects that alter the total energy and hinder thermalization. In a similar vein, we argue that at stronger coupling strengths $g/\sigma$, the Hamiltonian spectrum becomes dense and suppresses ETH-based thermalization. Bridging these models and identifying more concrete and identical themes remains an open question.

{\color{black} Finally, while the exact mechanism driving quantum chaos is not well understood as its analog in classical mechanics, it is widely known that non-integrable Hamiltonians display a subextensive number of conserved quantities that promote chaotic dynamics in the Hilbert space. This is similar to classical mechanics where an extensive number of conserved quantities hinder ergodicity and prevent thermalization. In our model, the long-range excitonic-disorder $\sigma$ breaks the integrability of the TC model, allowing for chaotic dynamics to become dominant due to the subextensive scaling of conserved quantities in the thermodynamic limit.}

\textit{Summary of the model and its salient features:} We study a chaotic TC model in which a disorder in exciton–exciton couplings and biexciton interactions induces quantum chaos and drives thermalization within the excitonic manifold. At small cavity–exciton coupling $g$, the cavity acts as a weak perturbation that effectively operates at noncontiguous time intervals due to very small Rabi oscillation frequencies. This effectively isolates the dynamics within the excitonic manifold in these intervals. As $g$ increases, the Rabi oscillation rate increases and the excitons no longer remain isolated. In this limit, we retrieve all the properties of the ordinary TC model.

{\color{black} \subsection{Heuristic Dependence on Scaling}

In the conventional TC model with uncoupled excitons, a symmetric bright state collectively exchanges energy with the cavity. This collective coupling scales as $\sqrt{N}$ in the number of excitons. In such a case, we write the population of the cavity as $P_{cav}(t) = sin^2(\sqrt{N}gt)$ and the model exhibits superradiance. This is still maintained on introducing an exchange coupling $J$ between the excitons. However, in the case of an all-to-all exchange coupling, two regimes emerge that influence the dynamics of the cavity, 

\begin{eqnarray}
    \label{eq:prob_cav_1}
    P_{cav}(t) = \frac{2g^2}{(J/2)^2 + 2g^2} \sin^2\left(\sqrt{(J/2)^2 + 2g^2}\cdot t \right)
\end{eqnarray}

When the light-matter coupling dominates the exciton exchange coupling $g \gg J$, the system exhibits superradiance in the same manner as in the conventional TC model. This can be seen by looking at Eq. \eqref{eq:prob_cav_1}. On the other hand, when $J \gg g$, the bright state and the cavity mode are very detuned. As a result, the population transfer from the bright state into the cavity is suppressed. This is also seen in Eq. \eqref{eq:prob_cav_1} where the amplitude prefactor becomes very small for large $J$. The cavity population scales as,

\begin{eqnarray}
    P_{cav}(t) &=& \frac{4Ng^2}{(N-1)^2J^2 + 4Ng^2} \nonumber \\
    &&\sin^2 \left(\sqrt{\left[\frac{(N-1)J}{2}\right]^2 + Ng^2} \cdot t\right)
\end{eqnarray}

Therefore, the ability of the model to exhibit significant superradiance depends on the ratio between the individual light-matter interaction coupling $g$ and the exciton exchange coupling $J$. \\

This description becomes more obscured as we introduce (i) biexciton interaction and (ii) randomize the exciton exchange and biexciton interactions. An analytical description of the scaling of $P_{cav}(t)$ with $N$ becomes challenging. However fortuitously, the randomization lends an interesting structure to the level spacing statistics of the energy spectrum in the form of a Wigner-Dyson distribution as seen in Section \ref{subsec:spectral_statistics}. Without relying on an analytic description, this metric serves as a suitable diagnostic metric for collective phenomena in our model. \\
    
For these reasons, (i) the individual light-matter couplings $g$ and their comparison with the exciton interactions $h_{ij}, u_{kl}$ become important, and (ii) since the collective exciton couplings $h_{ij}, u_{kl}$ also scale with $N$ and compete with the scaling of the light-matter interaction $g$, the effective parameter that tracks the two regimes we bring out in this model is $g/\sigma$. The thermodynamic limit only makes our regimes sharper as guaranteed by ETH.}

\section{Thermalization in Polaritons} \label{sec:therm_excitons}

Isolated and open quantum systems thermalize through fundamentally different mechanisms. In open systems, driving forces and dissipative dynamics obeying the detailed balance condition govern thermalization. However, in isolated systems, a non-integrable Hamiltonian induces quantum chaos and pushes subspaces to thermalize at long times. While the von Neumann entropy of the global state remains conserved under unitary evolution, it increases as a power law for subsystems within the global system \cite{kim2013ballistic, dymarsky2018subsystem}. This is because as correlations propagate across the Hilbert space, subsystems begin to entangle with each other, leading to a growth in their entropy \cite{deutsch2018eigenstate}. At long times, these subsystems become maximally entangled and attain a maximal entropy state. For a given energy, this maximized entropy state is the thermal state. In this manner, an isolated quantum system with a given energy observes localized thermalization in the thermodynamic limit. The ETH is a collection of ideas that describe thermalization in these settings. Given a non-integrable Hamiltonian, ETH describes these properties in the energy eigenbasis. This leads to two versions that are germane to our work:

\begin{enumerate}[label=(\roman*), leftmargin=10pt]
\item \textit{State-based interpretation:} The state-based version of ETH states that, at long times, the reduced density matrix $\rho_A \coloneq \text{Tr}_{\bar{A}}(\rho)$ of a subsystem $A$ takes the form \cite{garrison2018does},

\begin{eqnarray}
    \label{density_matrix_version}
    \rho_A = \mathrm{Tr}_{\bar{A}}\!\left[\frac{e^{-\beta H}}{\mathrm{Tr}\!\left(e^{-\beta H}\right)}\right]
\end{eqnarray}

where $H$ is any nonintegrable Hamiltonian. In other words, the density matrix of any subsystem $A$ that spans at most half the real space of a lattice converges to a thermal density matrix in the thermodynamic limit. Consequently, the entanglement entropy of this state approaches the thermal entropy under appropriate constraints \cite{magan2016random, dymarsky2018subsystem}. Since the global energy of the system remains conserved, the energy of the initial state sets a unique inverse temperature $\beta$ of the final thermal state as long as it lies in the region of finite energy density of the energy spectrum \cite{srednicki1994chaos, srednicki1996thermal}. As a result, every finite energy eigenstate of a nonintegrable Hamiltonian uniquely determines the thermal properties of its underlying subsystems.

Most quantum spectroscopic experiments measure properties of hybrid light-matter particles called polaritons, which are the energy eigenstates of prototypical models such as the TC model. As a result, every finite energy density eigenstate of the Hamiltonian in Eq. \eqref{full_hamiltonian} observes localized thermalization as long as quantum chaos dominates its dynamics. In Section \ref{sec:results}, we show that the weak coupling limit satisfies this condition.

\item \textit{Operator-based interpretation:} The operator-based version of ETH states that a certain class of observables display expectation values that are characteristic to thermal equilibrium at long times. Given such an observable $O$ expressed in the energy basis, this means that its diagonal matrix elements become smooth functions of energy, $O_{nn}=\mathcal{O}(E_n)$, and its offdiagonal elements vanish with small and rare subextensive fluctuations in the thermodynamic limit. From the state-version of ETH, we write the operator expectation value as,

\begin{eqnarray}
    \label{operator_version}
    \langle O\rangle_t = \text{Tr}[O\sigma^{th}_A]
\end{eqnarray}

where $\sigma^{th}_A$ is the right-hand side of Eq. \eqref{density_matrix_version}. From Eq. \eqref{operator_version}, we see that $\mathcal{O}(E_n)$ takes the form a Gibbs distribution and requires the class of operators to be local to some subspace $A$. As a result, these expectation values thermalize and consequently, their measurements in the energy basis achieve a steady-state at long times. In the chaotic TC model, this implies that expectation values of local observables, expressed in every finite energy density polariton state, thermalize and attain a steady-state.

\end{enumerate}

In summary, any initial state, with finite energy density in the chaotic TC model of Eq. \eqref{full_hamiltonian}, will thermalize under appropriate conditions. Since the density of states of such a many-body system is highest in the middle of the spectrum, a large class of initial states will display such thermal behavior. While the mechanism to identify the observables that exhibit thermal expectation values remains open, we show in Section \ref{sec:results} that a brute-force approach reveals that the cavity populations attain a steady state in the weak coupling limit.

\subsection{Thermalization Diagnostics}\label{subsec:therm_diagnostics}

We select an arbitrary initial state $\rho_0$ with energy $E_0 = \text{Tr}[H\rho_0]$, where $H$ is defined in Eq.~\eqref{full_hamiltonian}. Since unitary evolution of the system under $H$ conserves the total energy, we use the matching condition to calculate the inverse temperature $\beta^*$ of the corresponding thermal state \cite{garrison2018does},

\begin{eqnarray}
    \label{matching_condition}
    \text{Tr}[H\rho_0] = \text{Tr}[He^{-\beta^* H}/Z]
\end{eqnarray}

 where $e^{-\beta^*H}/Z$ is the thermal state and $Z=\text{Tr}[e^{-\beta^*H}]$ is the partition function. As a result, the choice of the initial state and its energy $E_0$ fixes the thermal state in isolated systems.

After propagating the initial state for time $t$, $\rho(t) = e^{-iHt}\rho_0 e^{iHt}$ $(\hbar = 1)$, we look at a reduced state of the system by tracing out its complement. Let $A$ be the subsystem corresponding to exciton or cavity modes such that it spans less than half the system; then $\rho_A\coloneq \text{Tr}_{\bar{A}}(\rho)$ obeys Eq.~\eqref{density_matrix_version} in the thermodynamic limit.

We employ the trace distance to quantify the distance between $\rho_A$ and $\sigma_A^{th}$,

\begin{eqnarray}
    \label{trace_distance}
    \text{D}[\rho_A(t),\sigma_A^{th}] = \frac{1}{2}\text{Tr}\sqrt{\left[(\rho_A(t) - \sigma_A^{th})^\dagger(\rho_A(t) - \sigma_A^{th})\right]}
\end{eqnarray}

where $\sigma_A^{th}$ is right-hand side of Eq. \eqref{density_matrix_version}. If two states $\rho_1, \rho_2$ have a trace distance equal to $1$, then they are perfectly orthogonal. On the other hand, if it is equal to $0$, they are perfectly identical. This metric is very sensitive to the results of the measurements performed on each of the quantum states. Even if the expectation of any observable $O$ produces small differences when measured in $\rho_1$ and $\rho_2$, their trace distance sensitively captures this and exhibits a large value.

Although Eq. \eqref{operator_version} delineates the properties of observables required for thermalization, it remains unclear how to exactly identify $O$ for a given Hamiltonian. The density matrix formalism serves as an upper bound that quantifies the number, size, and other properties of its underlying observables that thermalize. Heuristically, a systematic brute-force approach helps identify if certain observables converge to a steady state.

\section{Results and Discussion} \label{sec:results}

In this section, we numerically study the dynamics of our coupled exciton-cavity system governed by the Hamiltonian $H$ in Eq. \eqref{full_hamiltonian}. We begin by placing two photons in the optical microcavity and denote the initial state as $\ket{\psi(0)} = \ket{200..0}$. Here, the first site indexes the cavity and the rest index the excitons. We consider an excitonic lattice comprising $L=80$ sites that may be connected in a complex graph as shown in Fig. \ref{fig:exciton_network}. We arbitrarily index these excitons in the site basis in order to generalize our study. Since $\ket{\psi(0)}$ is not an eigenstate of $H$, it undergoes nontrivial dynamics under unitary evolution. We express the time-evolved state $\ket{\psi(t)}$ in the site basis as,

\begin{eqnarray}
    \label{expansion}
    \ket{\psi(t)} = \alpha_0(t)\ket{2;\textbf{0}} &+& \mathbb{P}_1 \left[\sum_{i=1}^L \alpha_i(t) \ket{1;1}\right] \nonumber \\
    &+& \mathbb{P}_2 \left[\sum_{j=L+1}^M \alpha_{j}(t) \ket{0; 11}\right]
\end{eqnarray}

where the first site indexes the cavity and the rest index the excitons. Additionally, $\mathbf{0}$ denotes the collective ground state of $L$ excitons, $\mathbb{P}_{1(2)}$ denotes all the permutations of a single (double) exciton(s) in its site basis, and $M$ denotes the number of permutations in $\mathbb{P}_2$. As a result, the dimension of the Hilbert space is $d=1+L+M=3241$, sufficient to mitigate finite-size effects from our simulations. Here, $\sum_{i=0}^d |\alpha_i(t)|^2=1$ preserves the norm of the final state.

To understand the ergodic properties of this model, we define the inverse participation ratio (IPR) as,

\begin{eqnarray}
    \label{ipr}
    \text{IPR} = \sum_{i=0}^d |\alpha_i(t)|^4
\end{eqnarray}

This quantifies the measure of localization in a many-body quantum state. For a $d-$dimensional Hilbert space, $\frac{1}{d} \leq \text{IPR} \leq 1$ where $\text{IPR} = \frac{1}{d}$ denotes a completely delocalized state and $\text{IPR} = 1$ denotes a completely localized state. Since we are interested in spectroscopically calculating the exciton coupling disorder $\sigma$, we tune $g/\sigma$ by tuning the coupling strength $g$ throughout this work. In doing so, we fix a sufficiently large $\sigma$ such that Eq. \eqref{ex_hamiltonian} is non-integrable and effectively captures ETH. This choice sets the energy scale for our simulations.

\begin{figure}
    \centering
    \includegraphics[width=0.95\linewidth]{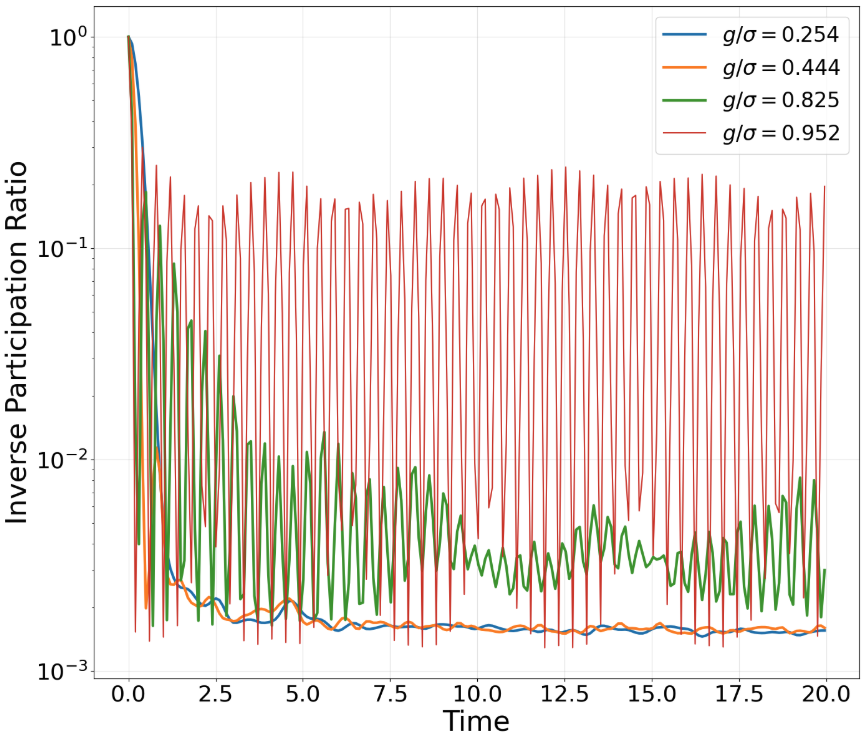}
    \caption{\textbf{Inverse Participation Ratio:} This plot shows the IPR in the two coupling regimes. At small coupling strengths $g$ (blue, yellow), the IPR decays and saturates $\approx \frac{1}{d}$. This indicates that all the excitations are delocalized throughout the Hilbert space and the system is ergodic. At stronger coupling strengths (green, red), the IPR oscillates in complete range, $\frac{1}{d} \leq \text{IPR}\leq 1$. This indicates that the system is non-ergodic during strong coupling strengths.}
    \label{fig:IPR}
\end{figure}

In Fig. \ref{fig:IPR}, we see that for small coupling strengths where $\sigma$ dominates $g$, the IPR decays and saturates nearly at $10^{-3}\approx \frac{1}{d}$. This indicates that the two photons present in the cavity at $t=0$ delocalize throughout the Hilbert space at long times and reach a steady state. Since the IPR is expressed in the site basis that represents the microstates of the system, its dynamics in this small coupling strength regime indicates that the system becomes ergodic. On the other hand, at larger coupling strengths where $g/\sigma \sim 1$, the IPR oscillates rapidly between $10^{-3}\text{ and } 10^{-1}$. With the same initial state where $\alpha_0(0)=1$, this shows that the excitations periodically localize within the cavity, and renders the system non-ergodic. These properties govern the thermalization ability of polariton systems, similarly studied in the interacting disordered Tavis-Cummings model from Ref. \cite{mattiotti2024multifractality}. 

Having looked at the ergodic properties and their corresponding regimes in our model, we now study its thermalization dynamics using the trace distance as described in Section \ref{subsec:therm_diagnostics}. To ensure that subsystems within our many-body state thermalize at long times, the energy of the initial state must lie far from the ground state \cite{garrison2018does}. With two photons in the cavity in the initial state, we see in Fig. \ref{fig:expectation_H} that the expectation of the Hamiltonian lies near the center of the energy spectrum. Therefore, $\ket{\psi(0)}$ serves as a viable state to observe localized thermalization at long times in the ergodic regime.

We begin by evolving $\rho(0)$ for some time $t$ to give the final state $\rho(t)=e^{-iHt}\rho(0)e^{iHt}$ where $\rho(0)=\ket{\psi(0)}\bra{\psi(0)}$. To understand the convergence of subsystems towards the thermal state, we take the partial trace of $\rho(t)$ over the complement of $A$ and obtain the reduced state $\rho_A(t) \coloneq \text{Tr}_{\bar{A}}[\rho(t)]$. In our simulations, we choose $A$ to span three arbitrarily chosen excitons so that the subspace dimension condition of ETH is satisfied. To ensure statistical robustness, we average over all possible and ETH-permissible spatial realizations of $A$ within the lattice, as well as over $5$ independent realizations of the disordered Hamiltonian, $H$. Finally, we measure the trace distance between $\rho_A(t)$ and its corresponding thermal state $\sigma_A^{th}$ as described in Eq. \eqref{trace_distance}.

\begin{figure}
    \centering
    \includegraphics[width=\linewidth]{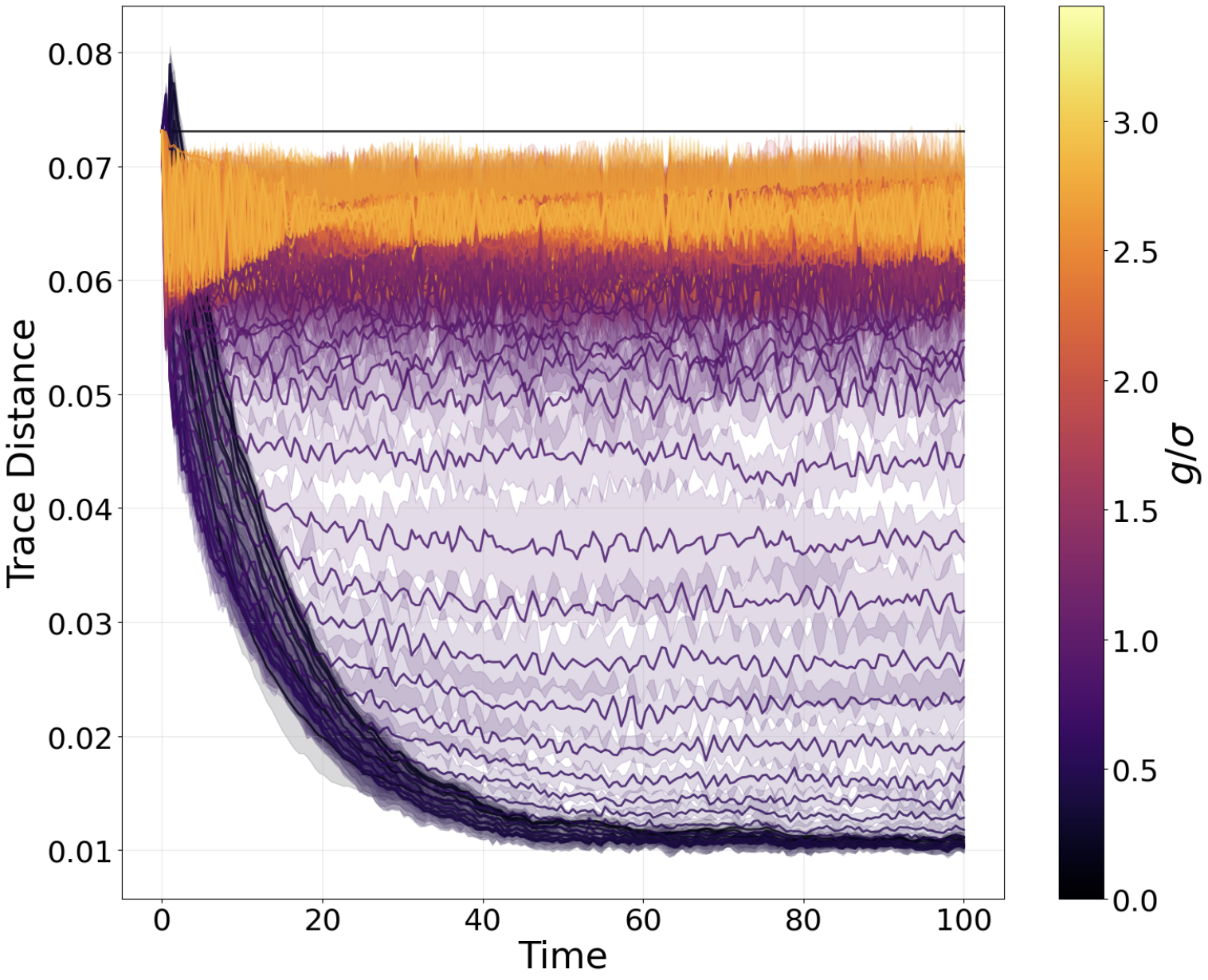}
    \caption{\textbf{Trace Distance Decay:} This figure plots the trace distance of a reduced system with time as a function of the ratio between the cavity-exciton coupling strength and thermalization rate. The initial state has 2 excitations in the cavity and $g$ is the coupling strength of the cavity with the excitonic lattice. We run this simulation over $5$ runs across multiple excitonic subspaces $A$ and observe that the mean trace distances numerically converge to a statistically stable state. The shaded regions represent one standard deviation away from the mean and their narrowness at weak $g/\sigma$ confirms that the subsystems consistently thermalize in the long-time limit.}
    \label{fig:loss_therm}
\end{figure}

In Fig. \ref{fig:loss_therm}, we observe that the solid lines representing the ensemble mean of the trace distances smoothly decay to zero in the ergodic regime $(g/\sigma \sim 0)$. In addition, the shaded regions that represent the ensemble fluctuations remain negligibly small at all times. These indicate that subsystems of the global isolated system thermalize at long times. We note that these trace distances do not return to their initial values at asymptotic times since we probe reduced systems within the global state. As a result, an absence of very low frequency Rabi oscillations is not an artifact of the finite time simulation, but a true statistical property of many-body quantum systems. In the non-ergodic regime $(g/\sigma \sim 1)$, the trace distances saturate at higher values with noticeably larger fluctuations as seen by the shaded regions. This indicates a gradual loss of thermalization driven by the competition between $g$ and $\sigma$. At much higher strengths $g/\sigma >1$, we see that the trace distances oscillate rapidly and observe larger fluctuations at all times. 

More physically, the coupling strength $g$ dictates the rate of Rabi oscillations between the cavity and the excitons, while the exciton coupling disorder $\sigma$ sets the timescale for the spread of excitations within the exciton manifold. As a result, at weak coupling strengths in the ergodic regime, the coupling disorder $\sigma$ dominates the Rabi oscillation frequency, leaving the excitations sufficient time to thermalize within the excitonic manifold. On the other hand, when the system is in the nonergodic regime at $g/\sigma \sim 1$, the Rabi oscillation frequencies compare with the exciton coupling disorder. As a result, the excitations in the excitons are perturbed before their subsystems begin to converge to a thermal state. At much higher values of $g/\sigma$, the interaction Hamiltonian in Eq. \eqref{cav_ex_coupling} dominates the excitonic Hamiltonian in Eq. \eqref{ex_hamiltonian} and effectively reduces $H$ to the conventional TC model. We see that running this simulation over $5$ runs across multiple excitonic subspaces $A$ remains sufficient for the mean trace distances to numerically converge to a statistically stable state. The shaded regions, which represent one standard deviation away from the mean, also remain narrow at weak $g/\sigma$ and confirm that the subsystems consistently thermalize in the long-time limit.

\begin{figure}[!htbp]
    \centering
    \includegraphics[width=\linewidth]{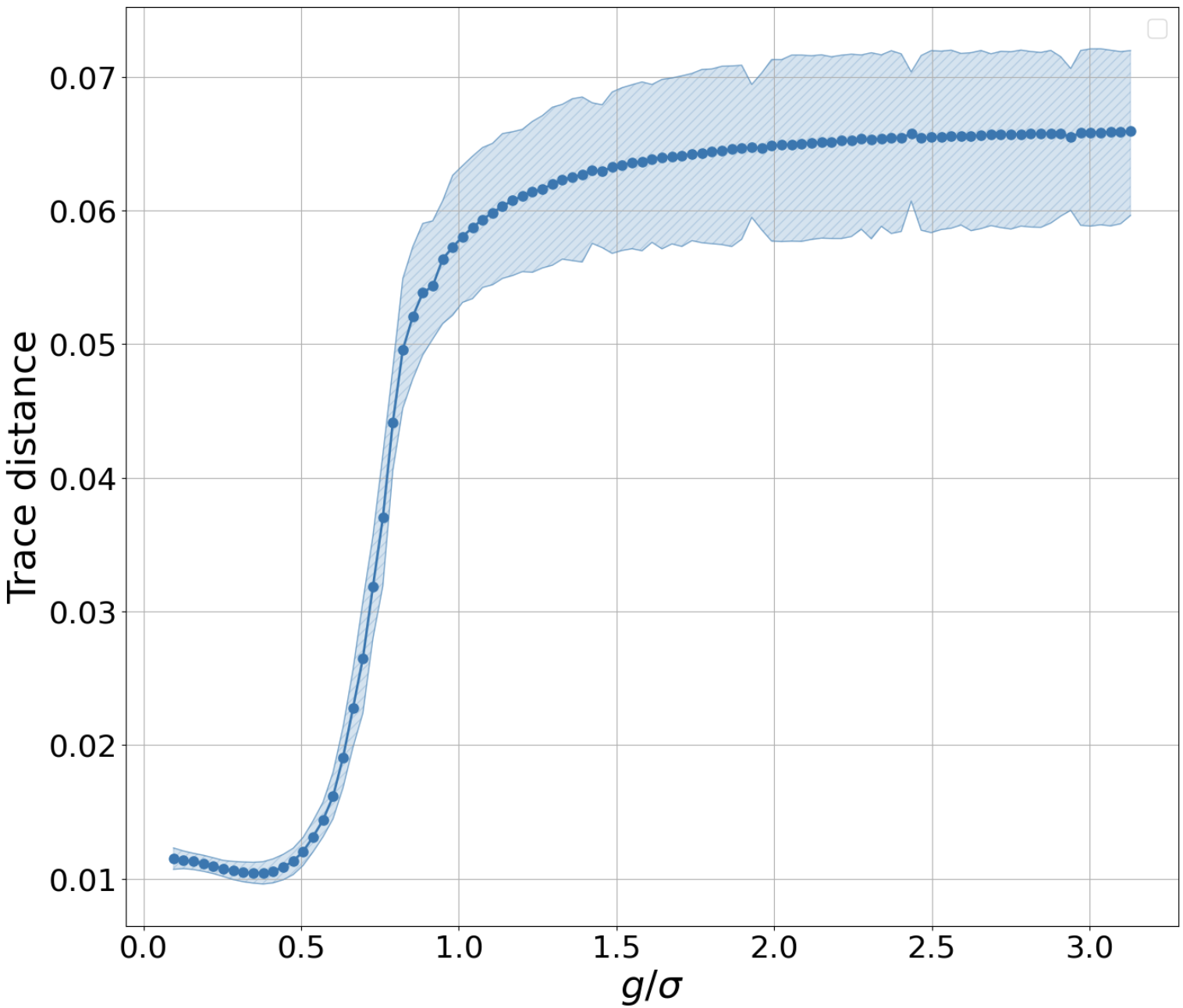}
    \caption{\textbf{Transition from a thermalizing to a non-thermalizing regime:} We see that at low polariton splittings, the trace distance of a reduced density matrix of excitons with its corresponding thermal state is close to zero and indicates a thermalization regime. As we increase the polariton splitting, it dominates the exciton coupling width and increases the final trace distance. This leads to a non-thermalizing regime close to $g/\sigma \sim 1$.}
    \label{fig:regimes_therm}
\end{figure}

To make this more coherent, we focus on the latter half of the time interval shown in Fig. \ref{fig:loss_therm} and take the moving average of the trace distances as the representative value in that interval. In Fig. \ref{fig:regimes_therm}, we see at weak coupling strengths, the representative points are close to zero with small fluctuations about the mean. From previous arguments on IPR, we label this region as the \textit{(i) ergodic regime of the chaotic Tavis-Cummings model} where the coupling disorder $\sigma$ dominates the cavity-exciton coupling strength $g$. As a result, the many-body exciton mixing dynamics promote ETH-based thermalization over TC-based Rabi oscillations. The level spacing statistics of the Hamiltonian in this regime resemble the WD distribution as shown in Fig. \ref{fig:level_spacings} and the model predominantly exhibits quantum chaos \cite{robnik2016fundamental}.

As we increase the coupling strength $g$, we see a sharp increase in the vicinity of $g/\sigma \sim 1$, indicating the transition away from the ergodic regime. This is because the strength of Rabi oscillations that remove excitations from the excitonic manifold compares with the strength of many-body mixing that promotes the spread of excitations within the excitonic manifold. 

At much higher coupling strengths, the representative points saturate at larger values of trace distances with large fluctuations about its mean. These two markers indicate that the system does not thermalize at long times and the TC-based Rabi oscillations dominate the ETH-based exciton thermalization. We label this region as the \textit{(ii) nonergodic regime of the chaotic Tavis-Cummings model} where the cavity-exciton coupling strength $g$ dominates the exciton coupling disorder $\sigma$, perturbing the exciton dynamics before they thermalize. In this regime, the polariton splitting, given by $\Delta=g\sqrt{N}$, is large and effectively masks the WD level spacings to reproduce the pure TC model. In Fig. \ref{fig:level_spacings}, we see that these statistics no longer resemble the WD distribution and the model does not exhibit quantum chaos. 

{\color{black} From the above discussion, we see that the cavity mode plays two roles in the different thermalization regimes we describe: (i) it acts as an adversary to the many-body thermalization due to ETH. When the cavity is strongly coupled to the excitons, the rate of exchange of excitations between the cavity and the material modes dominate the rate of delocalization within the excitons. More importantly, (ii) the cavity presents a great experimentally tunable knob that identifies the exciton coupling disorder of materials. By physically changing the experimental degrees of freedom of the cavity and scanning over multiple coupling strengths $g$, and adjusting the leakage of the cavity, one can directly understand the disorder $\sigma$. Moreover, at small coupling strengths $g$, the cavity ensures that excitations remain in the material and that they do not leak out before they thermalization. We elaborate more on this in Section \ref{sec:experimental_utility}.}

{\color{black} While the results so far represent the behavior of finite-sized systems due to classical computational capabilities, the theory of ETH ensures that they converge to more stable results in the large $N$ limit. More specifically, Fig. \ref{fig:loss_therm} would show a smoother and sharper convergence to a thermal state at small values of $g/\sigma$, and Fig. \ref{fig:regimes_therm} would show a sharper crossover from the thermalizing to the non-thermalizing regime with a sharper plateau in the latter. However, these results will not break down if the exciton interactions exhibit long-range order or are described by a $k-$local Hamiltonian where $k$ is small. In the latter, a series of Bogoliubov and Jordan-Wigner transformations can convert the Hamiltonian to integrable forms (such as the free-boson model) that prevent thermalization.}

{\color{black} The thermalization regime that we demonstrated so far arises primarily due to quantum chaos induced by the excitonic coupling disorder $\sigma$, This disorder makes the Hamiltonian non-integrable and promotes thermalization. In the same way, \citet{emary2003chaos} show that the Dicke model also presents a non-integrable Hamiltonian that exhibits quantum chaos. In the absence of the RWA, the Dicke Hamiltonian can be written as,

\begin{eqnarray}
    \label{eq:dicke_model}
    H/\hbar = \omega_c a^\dagger a + \omega_0 J_z + \frac{g'}{\sqrt{N}}J_x(a+a^\dagger)
\end{eqnarray}

In the $N\rightarrow \infty$ limit, there is quantum phase transition at $T=0$ at $g'_c = \frac{\omega_c \omega_0}{2}$ beyond which the model begins to exhibit quantum chaos marked by WD statistics in its energy level spacings. Thus, chaos arises when the system is pushed toward the superradiant phase ($g'>g'_c$). However, if we assume RWA, then the Dicke model neglects the counter-rotating terms, transforming it to a TC model, that make it integrable and inhibit quantum chaos. In our work, we introduce chaos in this model through a fundamentally different mechanism: by motivating the presence of a disorder in the excitonic interactions that promote quantum chaos despite invoking the RWA. Moreover, the Dicke model in Eq. \eqref{eq:dicke_model} becomes integrable in the $N\rightarrow \infty$ limit and precludes thermalization. In our chaotic TC model, the thermodynamic limit $N\rightarrow \infty$ improves final state thermal behavior as guaranteed by ETH. More succinctly,

\begin{itemize}
    \item Decreasing $N$ or $L$ (for example, $N\leq 1$, $L=40$) decreases the Hilbert space dimension, introduces finite-size effects and \textit{prevents the system from thermalizing}. However, increasing each of these quantities makes the thermalization dynamics more apparent and \textit{further converges our results}. 
    \item Increasing $g$ promotes Rabi oscillations that are characteristic of the TC model and prevents the system from thermalizing. On the other hand, decreasing $g$ strengthens the exciton-coupling disorder, weakens Rabi oscillations, and pushes the system to thermalize.
\end{itemize}

Although the origin of quantum chaos in these models may be different, they share a common diagnostic in the distribution of their energy level spacings. This is because quantum mechanics does not share the same mechanism of chaos as commonly seen in classical mechanics. As a result, while the origins of non-integrability in the two Hamiltonians may be different, their resulting chaotic dynamics are reflected in the energy level spacings.}

The results we showed in this section can also be observed in experiments. Since $g$ can be tuned, for example, by tuning the distance between the reflecting surfaces of the microcavity, these regimes can be easily realized in quantum optical experiments. In the next section, we propose experiments and observables that certify these underlying regimes.

\section{Experimental Utility} \label{sec:experimental_utility}

We propose that the theoretical model introduced in this work can be experimentally tested using EBS. In these experiments, a Type-I $\beta$-Barium Borate (BBO) crystal phase-matched for SPDC with the pump wavelength of $343$ nm \cite{kumar2026hyperspectral, malatesta2023optical} generates spectrally entangled biphoton states. We define this state on the input vacuum $\ket{0}_{in}$ as,

\begin{eqnarray}
    \ket{\psi}_{\text{in}} = \int d\omega_s d\omega_i  \mathcal{F}_{\text{in}}(\omega_s,\omega_i)b^\dagger_{\text{in}}(\omega_s) b^\dagger_{\text{in}}(\omega_i) |0\rangle_{\text{in}}
\end{eqnarray}

where $\omega_s, \omega_i$ are the conventional signal and idler channel frequencies, $b^\dagger_{\text{in}}(\omega_s), b^\dagger_{\text{in}}(\omega_i)$ are the photon input operators as described in Ref. \cite{gardiner2004quantum}, and $\mathcal{F}_{\text{in}}(\omega_s,\omega_i)$ is called the Joint Spectral Amplitude (JSA) that characterizes their spectral entanglement. When these biphotons are impinged on a microcavity, they populate it with two excitations. When these photons are in resonance with the bare cavity, $\omega_s = \omega_i = \omega_0$, this experiment produces the initial state with two excitations in the cavity as described in Section \ref{sec:results}. The energy of this state lies between the upper and lower polariton energies, as depicted in Fig. \ref{fig:expectation_H}.

At intermediate times, these excitations interact with the degrees of freedom of the underlying material according to Eq. \eqref{cav_ex_coupling} and realize different regimes depending on the experimentally tuned coupling strength $g$. At some later time, the cavity modes are converted back to output biphotons and are incident on a detector. Our previous work \cite{dambal2025quantum} constructs a theoretical framework that connects the input variables at $t\rightarrow-\infty$ and the output variables at $t\rightarrow +\infty$ to determine how material correlations leave imprints on the output JSA.

These experiments use intensity interferometry \cite{boal1990intensity} to probe the output photon statistics in the photodetectors. Two such detectors are placed based on the output photon wavevectors and measure biphoton coincidence detection rates in a given time interval. These coincidences measure the second-order correlation function $g^{(2)}(t+\tau)$ defined as \cite{carmichael1993open},

\begin{eqnarray}
    g^{(2)}(t+\tau) = \frac{\langle a^\dagger(t) a^\dagger(t+\tau) a(t+\tau) a(t) \rangle}{\langle a^\dagger(t) a(t) \rangle \langle a^\dagger(t+\tau) a(t+\tau) \rangle}
\end{eqnarray}


where $\tau$ is the time-delay between two subsequent photon detections. At zero time delay ($\tau=0$), this quantity measures the photon bunching and antibunching properties and indicates the classical and quantum nature of the source of light. At finite time delay ($\tau \neq 0$), this quantity measures the correlations in the photon detection signal and reveals quantum correlations of the light source. {\color{black} Such photon correlation measurements of the TC model are extensively studied in literature \cite{gu2020manipulating, carnio2021optimization, stegmann2022higher}}

\begin{figure}
    \centering
    \includegraphics[width=\linewidth]{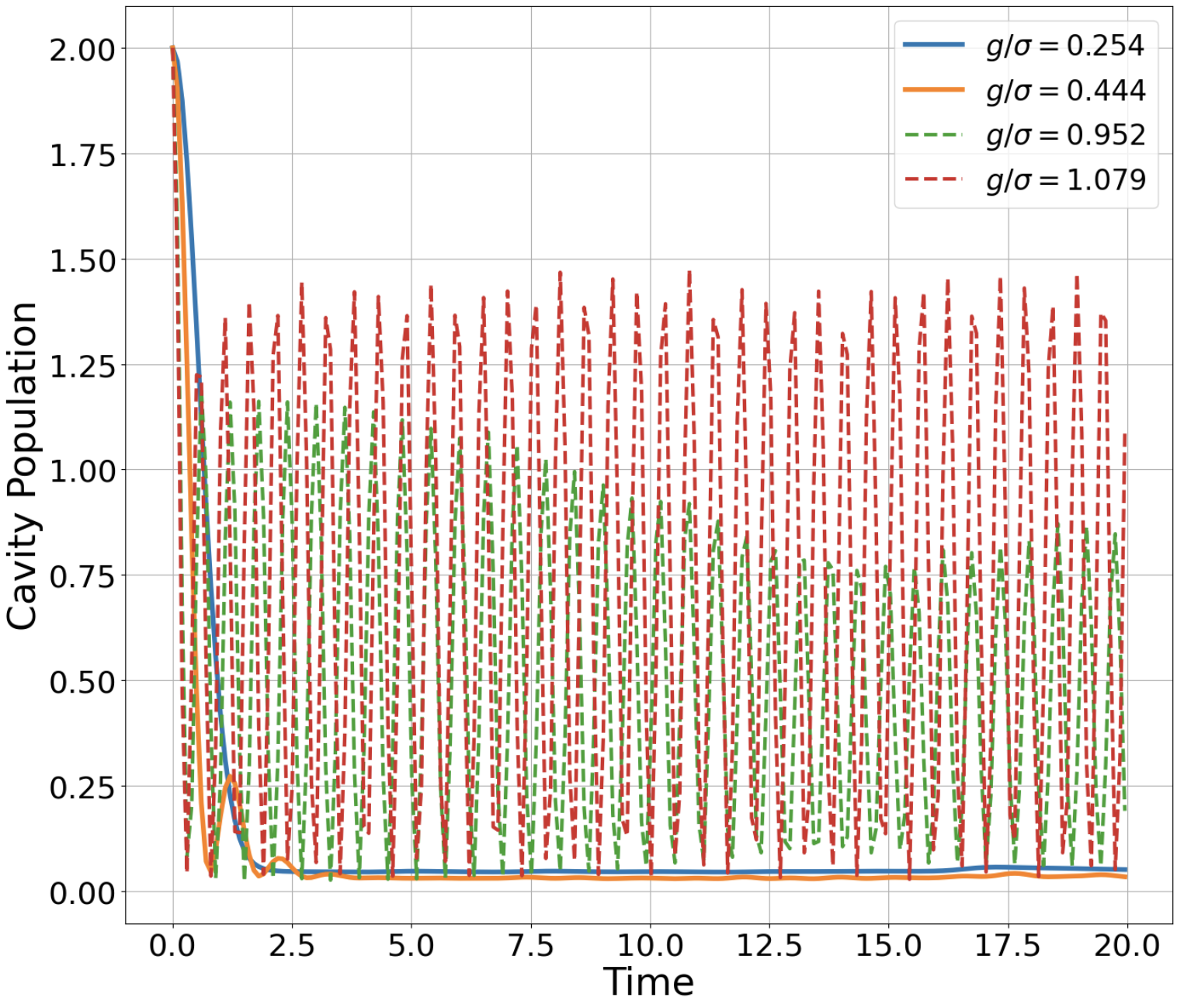}
    \caption{\textbf{Cavity population dynamics:} At small coupling strengths $g$ (ergodic regime, solid lines), the cavity populations decay and attain a steady state after a short time. At stronger coupling strengths (non-ergodic regime, dotted lines), the cavity populations oscillate and do not decay at long times.}
    \label{fig:cavity_pop}
\end{figure}

In our theoretical model, the cavity populations that convert into the output photons serve as the source of light on the photodetectors. The quantum regression theorem \cite{carmichael1993open} then implies that the correlation time of the cavity populations is proportional to the correlation time observed in $g^{(2)}(t+\tau)$ measurements. As a result, understanding the cavity population dynamics in the thermalizing and non-thermalizing regimes provides a direct insight into the differing correlation times observed in $g^{(2)}(t+\tau)$ measurements.

In Fig. \ref{fig:cavity_pop}, we define the cavity populations $n_c(t) \coloneq a^\dagger(t)a(t) = 1p_1(t) + 2p_2(t)$ and plot them for the different coupling strengths $g/\sigma$. Here $p_1(t) = \sum_{i=1}^L |\alpha_i(t)|^2, p_2(t)=|\alpha_0(t)|^2$ (c.f.~Eq.~\eqref{expansion}) are the probabilities of finding a single and double excitation in the cavity. We see that in the ergodic regime described in Section \ref{sec:results} and Fig. \ref{fig:IPR}, the cavity populations decay at long times. Since the system is ergodic, this indicates that the initial excitations from the cavity delocalize throughout the Hilbert space and attain a steady state. Moreover, the cavity population describes a local operator defined on the site basis. In Section \ref{sec:therm_excitons}, we mentioned that while operators must be local in order to thermalize, only a brute-force approach to identifying them reveals these operators. Figure \ref{fig:cavity_pop} shows that $n_c(t)$ is one such operator that obeys the operator-version of ETH. 

In the nonergodic regime, we see that the cavity populations oscillate and fail to attain a steady state at long times. In this regime, we retrieve the collective dynamics characteristic of the pure TC model. A closer look at these time signals reveals that the correlation times markedly change in going from the ergodic to nonergodic regimes. To understand this in more detail, we calculate the population fluctuations $\delta n_c(t) = n_c (t) - \bar{n}_c$ where $\bar{n}_c=\frac{1}{T}\int_0^T n_c(t)dt$ is the mean population of the cavity. We then define the autocorrelation function

\begin{eqnarray}
    C(\tau) \coloneq \int_0^T \delta n_c(t) \delta n_c(t+\tau)dt
\end{eqnarray}

and calculate $C(\tau)$ using the Wiener-Khinchin theorem

\begin{eqnarray}
    C(\tau) = \mathcal{F}^{-1}[S(\omega)](\tau)
\end{eqnarray}

where $S(\omega)=|\mathcal{F}[\delta n_c(t)]|^2$ is the power spectral density and $\mathcal{F}[\cdot]$ is the Fourier transform. This gives the correlation time $\tau_c$ of the time-signal

\begin{eqnarray}
    \tau_c = \int_0^\infty \left|\frac{C(\tau)}{C(0)}\right|d\tau
\end{eqnarray}

\begin{figure}
    \centering
    \includegraphics[width=0.95\linewidth]{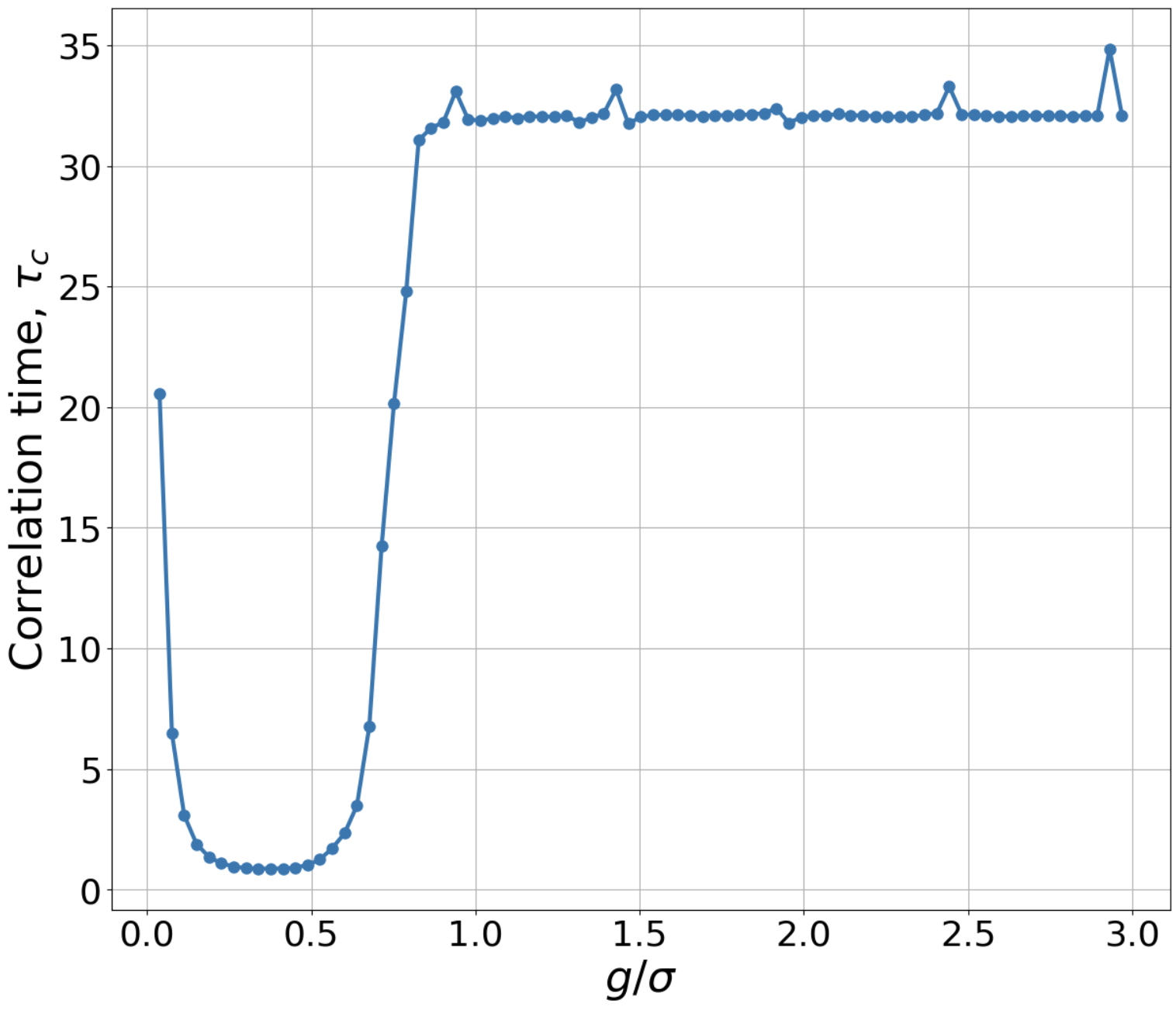}
    \caption{\textbf{Correlation time of cavity populations:} We see that if the polaritons are in the ergodic regime, the output photon statistics have a short correlation time. However, in the nonergodic regime, the correlation time increases until it saturates. This serves as a direct experimental observable that informs the underlying exciton coupling disorder $\sigma$. }
    \label{fig:correlation_time}
\end{figure}

In Fig. \ref{fig:correlation_time}, we see that in the ergodic regime, the cavity populations exhibit short correlation times and reflect the rapidly decaying nature of its corresponding time signal. At $g/\sigma \sim 1$, the correlation time observes a sharp spike and indicates that the cavity perturbs the system out of ergodicity. In the non-ergodic regime, we see that the correlation times saturate at a higher value, reflecting the oscillatory nature of its corresponding time signal. The initial decay in the correlation time indicates the incipient filling of the excitons from the initial cavity state.

In our theoretical model, the exciton-coupling disorder $\sigma$ sets the energy scale of the system. Since this is characteristic to a given material and is the parameter of interest to quantum spectroscopy, we experimentally tune $g$ and identify $g/\sigma \sim 1$ by observing the change in the $g^{(2)}(t+\tau)$ correlation times. To exemplify our discussion, we refer to Ref.~\cite{quiros2026resolving}, where the authors conduct an experiment that looks at exciton-polariton correlations in optical microcavities and estimate the light-matter coupling strength to be $g \approx 80$~meV. Using this as a baseline and our results from Fig. \ref{fig:correlation_time}, we propose that if one tunes the light-matter coupling $g$ and finds that at $g \approx 80$~meV, the $g^{(2)}(t+\tau)$ measurements exhibit a short correlation time on the order of $\tau_c \approx 0.53767$~ps, then the underlying excitonic-coupling disorder $\sigma \gtrsim 80$~meV. On the other hand, if the correlation time is $\tau_c \approx 3.85023$~ps, then $\sigma \lesssim 80$~meV. In regions where $\tau_c$ observes a transition between these two correlation times, $\sigma \sim g \sim 80$~meV. While the specific magnitudes of $\tau_c$ depend on the properties of the underlying material, the trend in the change of $\tau_c$ will hold for all materials with a nontrivial Hamiltonian from Eq. \eqref{ex_hamiltonian}.


These results serve as a useful reference for intensity interferometry experiments to control the cavity-material coupling strength $g$, measure the correlation times of $g^{(2)}(t+\tau)$ observables and leverage the underlying ergodic regimes to extract the exciton-coupling disorder $\sigma$ of a given material. {\color{black}To ensure that the phenomena we propose in this work is observed experimentally, the cavity leakage rate must be lesser than the thermalization timescale. In the absence of a signal, one can experimentally increase the cavity photon lifetime and measure the relevant observables.} Put together, our results describe a framework that uses specific experimentally controllable parameters to understand the many-body correlations of materials at the single-excitation level using quantum light spectroscopy. 

\section{Conclusion and Future Directions}


In this work, we studied thermalization in a chaotic Tavis–Cummings model and showed that the interplay between exciton–exciton coupling disorder and cavity-mediated interactions gives rise to two distinct dynamical regimes. In the weak-coupling limit, where disorder dominates, the system exhibits quantum chaos characterized by Wigner–Dyson statistics and satisfies the conditions of ETH. In this regime, excitations delocalize across the Hilbert space, and local observables, such as cavity populations, relax to steady-state values. In contrast, at strong cavity–exciton coupling, coherent Rabi oscillations dominate the dynamics, suppressing ergodicity and preventing thermalization. The system instead retains memory of its initial state, reflected in persistent oscillations and long-lived correlations.

A key result of this work is that these regimes leave clear, measurable signatures in experimentally accessible quantities. In particular, we showed that the correlation time of the output photon statistics, obtained through $g^{(2)}(t+\tau)$ measurements in EBS, directly tracks the underlying thermalization dynamics. Short correlation times correspond to ergodic, thermalizing behavior, while long correlation times signal non-ergodic dynamics. This provides a practical route to infer the exciton coupling disorder $\sigma$ by tuning the cavity coupling $g$, thereby connecting many-body quantum dynamics with experimentally controllable parameters.

More broadly, our results highlight how the inevitable many-body mixing within the excitonic manifold can be leveraged to extract microscopic material properties using quantum light. In this sense, the chaotic Tavis–Cummings model serves as a bridge between quantum statistical mechanics and quantum spectroscopy, where concepts such as ETH acquire direct experimental relevance.

\textit{Future Directions:} A more detailed understanding of how spectral statistics reorganize across the ergodic–non-ergodic transition could provide deeper insight into the mechanisms that suppress thermalization at strong coupling. An analytical description of this transition would further clarify whether it can be viewed as a sharp phase transition or a crossover. Additionally, since thermalization occurs in finite-energy-density polaritonic eigenstates, exploring the structure and dynamics of these eigenstates may reveal properties of these elusive dark states. Finally, dephasing has been shown to accelerate thermalization in quantum systems \cite{dambal2025noise}. Incorporating this may extend these results to more realistic experimental settings and potentially broaden the parameter regimes where thermalization can be observed.

Taken together, this work establishes a framework in which many-body thermalization, quantum chaos, and photonic observables are tightly linked, opening a pathway for investigating complex material correlations using quantum optical platforms.



\section*{Author Contributions}

S.D. conceived the project, developed the theoretical framework, performed numerical simulations, and drafted the manuscript. E.R.B. supervised the project, analyzed results, and contributed to discussions and manuscript revisions. All authors reviewed and approved the final version of the manuscript.

\begin{dataavailability}
The data supporting the findings of this study are publicly available on the \href{https://github.com/BittnerTheoryGroup/Thermalized_Polaritons}{Bittner Theory Group} GitHub repository.
\end{dataavailability}

\acknowledgments
The authors thank Pavan Hosur and Ajay Ram Srimath Kandada for useful discussions.
The work at the University of Houston was supported by the National Science Foundation under CHE-2404788 and the Robert A. Welch Foundation (E-1337).

\bibliography{bib-local}

\end{document}